\documentclass[preprint,aps,floatfix]{revtex4-1}
\usepackage{hyperref}
\everymath={\displaystyle}
\usepackage{young}
\usepackage{rotating}
\usepackage{longtable}
\def\1{\mbox{l\hspace{-0.53em}1}}

\usepackage{multirow}
\newlength{\AccoHaut}

\begin{document}
\title{Exploring the spectrum of the hidden charm strange pentaquark in the SU(4) 
version of the flavor-spin model}
\author{Fl. Stancu\thanks{e-mail : fstancu@ulg.ac.be}}
\address{Universit\'{e} de Li\`ege, Institut de Physique B.5, Sart Tilman,
B-4000 Li\`ege 1, Belgium}

\date{\today}
\everymath={\displaystyle}

\begin{abstract}
We study the spectrum of the isoscalar pentaquark $udsc\overline{c}$,
of either positive or  negative parity,  
in a constituent quark model with linear confinement and a  flavor-spin hyperfine interaction
previously extended to SU(4) and used to describe the spectrum of the $uudc\overline{c}$ pentaquarks observed at LHCb in 2019.
For positive parity we make a distinction between the case where one unit of angular momentum 
is located in the subsystem of four quarks and the case where the angular momentum is located 
in the relative motion between a ground state four-quark subsystem and the antiquark. 
The novelty is that we introduce the coupling between different
flavor states, due to the breaking of exact SU(4)-flavor symmetry of the Hamiltonian model, 
both for  positive and negative parity states.
An important consequence is that the lowest state, located at 4404 MeV, has quantum numbers $J^P$ = $1/2^-$
while without coupling the lowest  state has $J^P$ = $1/2^+$ or $3/2^+$.
\end{abstract}

\maketitle

\vspace{1cm}

\section{Introduction}

The 2019 LHCb observation of the narrow structures  $P^+_c(4312)$, $P^+_c(4440)$ and $P^+_c(4457)$
in the $\Lambda^0_b \rightarrow J/\psi K^- p$ decay \cite{Aaij:2019vzc} 
has given a new impetus to the study of hidden charm pentaquarks. The  $J/\psi p$ component 
suggested that the pentaquark wave functions should have the flavor content  $uudc\overline{c}$. 

Although observed in the $J/\psi p$ channel,
the proximity of the mass of the $P^+_c(4312)$ to the $\Sigma^+_c \overline{D}^0$ threshold  (4318 MeV)
and of the masses of  $P^+_c(4440)$ and $P^+_c(4457)$ to the  $\Sigma^+_c \overline{D}^{*0}$ threshold (4460 MeV),
favored  their interpretation as molecular S-wave of the   $\Sigma^+_c + \overline{D}^0$ 
and $\Sigma^+_c + \overline{D}^{*0}$  systems respectively
 \cite{Guo:2019kdc,Guo:2019fdo,Xiao:2019mst,Xiao:2019aya,Shimizu:2019ptd,Lin:2019qiv,
Liu:2019tjn,Meng:2019ilv,Wu:2019rog,Valderrama:2019chc,Du:2019pij,Wang:2019spc,Xu:2020gjl,Chen:2020pac}.
In such an interpretation, 
the binding arises via meson exchanges between point particles and in the elastic channel 
all resonances acquire a negative parity. 
However, if one introduces the coupling of the $\Sigma^+_c \overline{D}^{*0}$ and the $\Lambda_c(2595) \overline{D}$ channels,
due to the very close proximity of their thresholds, one 
obtains $J^P(4440) = {3/2}^-$  and   $J^P(4457) = {1/2}^+$ respectively \cite{Burns:2019iih}. 

A more general point of view has been adopted in Ref. \cite{Fernandez-Ramirez:2019koa} where the $P_c(4312)$ signal
was analyzed by using some general principles of the S-matrix theory. In this way it was concluded that $P_c(4312)$
is more likely a virtual (unbound) molecular state.

The 2019 LHCb pentaquarks have also been analyzed in compact pentaquark models based on the chromomagnetic interaction 
of the one gluon exchange model,
with quark/antiquark correlations \cite{Ali:2019npk} or without correlations \cite{Weng:2019ynv,Cheng:2019obk}.
In both cases the lowest state has negative parity.

Presently, the spin and parity of the narrow structures  $P^+_c(4312)$, $P^+_c(4440)$ and $P^+_c(4457)$
remains to be established experimentally.

Anticipating new experiments,
the 2019 LHCb successful observation stimulated interest in the theoretical study of analogue pentaquarks 
in particular of the 
hidden charm pentaquarks with strangeness, the $udsc \overline{c}$ system. For example, in Ref. \cite{Xiao:2019gjd}
it has been analyzed in the framework of a molecular scenario with heavy quark symmetry constraints and
in Ref. \cite{Wang:2019nvm} within the chiral effective theory where the short range contact interaction, 
the long range one-pion-exchange and the intermediate range two-pion-exchange interaction were included.
In Ref. \cite{Ferretti:2020ewe} the hidden charm pentaquarks with strangeness have been considered in the 
hadrocharmonium model.

Predictions for the isoscalar $udsc\overline{c}$ pentaquark have already been made previously. 
In Ref. \cite{Yuan:2012wz} the  spectrum of the $udsc\overline{c}$ pentaquark  was studied in the compact pentaquark picture
in a quark model with either the chromomagnetic, the flavor-spin or the instanton induced interaction. In all cases
it was found that the lowest state has the spin-parity $J^P = 1/2^-$. In Ref. \cite{Park:2018oib} 
the stability of several pentaquark systems  has been analyzed in a constituent quark model with
a simple chromomagnetic interaction, and  the $udsc\overline{c}$ pentaquark has been found among the 
most stable ones.

In an SU(4) classification of pentaquarks and its decomposition in SU(3) submultiplets, by selecting 
those with the charm quantum number $C$ = 0, 
one finds the $udsc\overline{c}$ pentaquark as a member of either an octet with isospin I = 0, 1 
or as a member of a decuplet with isospin I = 1. These SU(3) submultiplets belong to the 
$[421]$ irreducible representation of SU(4) of dimension 140. 
The members of the irreducible representation denoted by 140 can have a spin value of either 1/2 or 3/2
\cite{Wu:2004wg,Ortiz-Pacheco:2018ccl}.

The hidden charm pentaquarks having a strange quark are presently unknown. In principle they can be produced
and observed, for example, in the study of the  $\Xi^{-}_b \rightarrow J/\psi \Lambda K^-$ reaction  \cite{Xiao:2019gjd}
or in the decay of $\Lambda_b$ into $ J/\psi \Lambda K^0$ \cite{Lu:2016roh}.
Their discovery  would require much more data 
relative to the non-strange hidden charm pentaquarks observed at LHCb \cite{Ali:2019clg}.
If discovered they may possibly distinguish between the various theoretical pictures.

Here we explore the spectrum of the pentaquark  $udsc \overline{c}$  within a quark model 
\cite{Glozman:1995fu}, which has a  flavor dependent hyperfine interaction.
The hyperfine splitting in
hadrons is due to the short-range part of the Goldstone boson exchange
interaction between quarks. The merit of the flavor-spin (FS) model   
is that it reproduces the correct ordering of positive and negative parity
states of both nonstrange and strange baryons \cite{Glozman:1995fu,Glozman:1996wq,Glozman:1997jy}   
in contrast to the one gluon exchange (OGE) model.
However, it cannot explain the hyperfine splitting in mesons, because
it does not explicitly contain a quark-antiquark interaction.

It is therefore useful to compare  the spectrum of hidden charm nonstrange 
and hidden charm strange pentaquarks within the same model.

In a previous work  \cite {Stancu:2019qga}  
the  model of Ref. \cite{Glozman:1995fu} has been generalized from SU(3) to SU(4) in order to incorporate the
charm quark. The extension has been made in the spirit of the phenomenological approach
of Ref. \cite{Glozman:1995xy} where, in addition to Goldstone bosons of 
the hidden approximate chiral symmetry of QCD,  the flavor exchange interaction was augmented  
by an additional exchange of $D$ mesons between $u, d$ and $c$ quarks and 
of $D_s$ mesons between $s$ and $c$ quarks. The model provided a satisfactory description of 
the heavy flavor baryons.

The extended SU(4) flavor-spin model has been applied to the study 
of  $uudc \overline{c}$  pentaquarks.
Presently we study  the pentaquarks of structure $uds c \bar c$ in the same framework
considering both positive and negative parities.

The parity of the
pentaquark is given by P$\ = {\left({-}\right)}^{{\ell\ +\ 1}}$, where  $\ell$ is the orbital angular momentum.
As shown in Ref. \cite {Stancu:2019qga},  there are two ways to introduce orbital excitations.
For the lowest positive parity states
one way is to  introduce an angular momentum  $\ell$ = 1 in the internal motion of the four-quark subsystem
and the other is to introduce an unit of  angular momentum 
in the relative motion between a ground state four-quark subsystem and the antiquark.
According to the Pauli principle, in the first case  the four-quark subsystem  must be in a state of
orbital symmetry ${\left[{31}\right]}_{O}$. In the second case the four-quark subsystem 
is in the ground state ${\left[{4}\right]}_{O}$.  

In Ref. \cite{Stancu:2019qga}, in  the context of a schematic 
flavor-spin interaction, $\it i. e.$ exact SU(4) symmetry,  it was shown that the lowest pentaquark state has a positive parity
with the orbital excitation in the internal motion of the four-quark subsystem.   
Although the kinetic energy of such
a state is higher than that of the totally symmetric ${\left[{4}\right]}_{O}$
state of negative parity, the flavor-spin interaction overcomes this excess and generates a lower 
eigenvalue for the   ${\left[{31}\right]}_{O}$  state  with an  $s^3p$ configuration  than for 
${\left[{4}\right]}_{O}$  with an $s^4$ configuration.

In the exact SU(4) limit the strength of the interaction  is the same 
for all pairs, independent of the quark masses, and it is a constant as a function of the relative distance 
between the interacting quarks. The model Hamiltonian introduced in the next section
breaks the SU(4)-flavor symmetry through the quark masses and the radial dependence of the interaction
potential. We calculate the masses of the lowest positive and negative parity states of 
the  pentaquarks of structure $uds c \bar c$ considering states with flavor 
symmetry $[22]_F$, $[31]_F$ and $[211]_F$. The  SU(4)-flavor symmetry breaking implies the mixing
of wave functions containing $[31]_F$ and $[211]_F$ parts. It is shown that this mixing affects the ordering of positive
and negative parity states and that the lowest state $uds c \bar c$ pentaquark has quantum numbers $J^P$ = $1/2^-$.

The paper is organized as follows. In Sec. \ref{Hamiltonian} we introduce the 
model Hamiltonian and  the two-body matrix elements of the FS interaction 
corresponding to SU(4). Sec. \ref{coord} describes the orbital part of the 
four quark subsystem constructed to be translationally invariant both for positive 
and negative parity states.
Sections \ref{ke},\ref{linearcon} and \ref{FSint}
summarize analytic formulas.
Sec. \ref{numerical} contains the numerical results for the spectrum 
and a comparison with relevant previous studies of hidden charm strange pentaquarks. 
The last Section is devoted to
conclusions. 
Appendix 
\ref{Casimir}  is a reminder of useful group theory formulae for SU(n).
Appendix 
\ref{baryons} exhibits a variational solution for the baryon masses relevant for the 
present study.
In 
Appendix 
\ref{flavor4} we present explicit forms of the flavor states 
of content $udsc$ in the Young-Yamanouchi-Rutherford basis, for specific irreducible representations $[{f}]_F$. 


\section{The Hamiltonian}\label{Hamiltonian}
Here we closely follow the description of the model as presented in Ref. \cite{Stancu:2019qga}.
The parameters required by the incorporation of  the strange quark were  added.
  
The FS model Hamiltonian has the general form \cite{Glozman:1995fu}
\begin{eqnarray}
H &=& \sum_i m_i + \sum_i\frac{{\vec p}_{i}^2}{2m_i} 
- \frac {(\sum_i {\vec p}_{i})^2}{2\sum_i m_i} + \sum_{i<j} V_{\text{conf}}(r_{ij}) \nonumber\\
&+& \sum_{i<j} V_\chi(r_{ij}),
\label{ham}
\end{eqnarray}
with $m_i$ and ${\vec p}_{i}$ 
denoting the quark masses and momenta respectively
and $r_{ij}$ the distance between the interacting quarks $i$ and $j$. 
The Hamiltonian contains the internal kinetic energy and the linear confining interaction 
\begin{equation}
 V_{\text{conf}}(r_{ij}) = -\frac{3}{8}\lambda_{i}^{c}\cdot\lambda_{j}^{c} \, C\, r_{ij} \, .
\label{conf}
\end{equation}
The hyperfine part $V_\chi(r_{ij})$ has a flavor-spin structure extended to SU(4) in Ref. \cite{Stancu:2019qga}.
One has
\begin{eqnarray}
V_\chi(r_{ij})
&=&
\left\{\sum_{F=1}^3 V_{\pi}(r_{ij}) \lambda_i^F \lambda_j^F \right. +  \sum_{F=4}^7 V_{K}(r_{ij}) \lambda_i^F \lambda_j^F 
\nonumber \\
&+& \left.  V_{\eta}(r_{ij}) \lambda_i^8 \lambda_j^8 
+V_{\eta^{\prime}}(r_{ij}) \lambda_i^0 \lambda_j^0\right. 
\nonumber \\
&+& \left. \sum_{F=9}^{12} V_{D}(r_{ij}) \lambda_i^F \lambda_j^F\right.     
+ \left. \sum_{F=13}^{14} V_{D_s}(r_{ij}) \lambda_i^F \lambda_j^F \right.
\nonumber \\
&+& \left. V_{\eta_c}(r_{ij}) \lambda_i^{15} \lambda_j^{15} \right\}
\vec\sigma_i\cdot\vec\sigma_j, 
\label{VCHI}
\end{eqnarray}
\noindent
with the SU(4) generators $\lambda^F_i$ ($F$ = 1,2,...,15)
and
$\lambda^0_i = \sqrt{2/3}~{\bf 1}$, where $\bf 1$ is the $4\times4$ unit
matrix.

In the SU(4) version
the interaction (\ref{VCHI})
contains $\gamma = \pi, K, \eta, D, D_s, \eta_c$ and $\eta '$
meson-exchange terms.  Every $V_{\gamma} (r_{ij})$ is
a sum of two distinct contributions: a Yukawa-type potential containing
the mass of the exchanged meson and a short-range contribution of opposite
sign, the role of which is crucial in baryon spectroscopy. 
For a given meson $\gamma$ the meson exchange potential is
\begin{eqnarray}\label{radialform}
V_\gamma (r) &=&
\frac{g_\gamma^2}{4\pi}\frac{1}{12m_i m_j}
\{\theta(r-r_0)\mu_\gamma^2\frac{e^{-\mu_\gamma r}}{ r} \nonumber\\
&-& \frac {4}{\sqrt {\pi}}
\alpha^3 \exp(-\alpha^2(r-r_0)^2)\}
\end{eqnarray}
In the present calculations we use the
parameters of Ref. \cite{Glozman:1996wq} to which we add the $\mu_{D}$ and the $\mu_{{D}_s}$ masses and
the coupling constants $\frac{g_{Dq}^2}{4\pi}$ and $\frac{g_{D_sq}^2}{4\pi}$. These are 
$$\frac{g_{\pi q}^2}{4\pi} = \frac{g_{\eta q}^2}{4\pi} =
\frac{g_{Dq}^2}{4\pi}= \frac{g_{D_sq}^2}{4\pi}=          0.67,\,\,
\frac{g_{\eta ' q}^2}{4\pi} = 1.206 , $$
$$r_0 = 0.43 \, \mbox{fm}, ~\alpha = 2.91 \, \mbox{fm}^{-1},~~
 C= 0.474 \, \mbox{fm}^{-2}, \, $$
$$ \mu_{\pi} = 139 \, \mbox{ MeV},~ \mu_{\eta} = 547 \,\mbox{ MeV},~
\mu_{\eta'} = 958 \, \mbox{ MeV},~
 \mu_{K} = 495 \, \mbox{ MeV},\,$$
$$ \mu_{D} = 1867 \, \mbox{ MeV},~ \mu_{{D}_s} = 1968  \mbox{ MeV}. $$
The meson masses correspond to the experimental values from  the  Particle Data Group \cite{Tanabashi:2018oca}.
As discussed 
in the following, we ignore the $\eta_c$-exchange.

The model of Ref. \cite{Glozman:1996wq} has previously been used to study the stability of
open flavor tetraquarks \cite{Pepin:1996id} and open flavor pentaquarks \cite{Genovese:1997tm,Stancu:1998sm}.
Accordingly,
for the quark masses we take the values determined variationally in 
Refs. \cite{Pepin:1996id,Genovese:1997tm} 
\begin{equation}\label{quarkmass}
 m_{u,d} = 340 \, \mbox{ MeV},~ m_s = 440 \,\mbox{ MeV}, ~ m_c = 1350 \, \mbox{ MeV}.
\end{equation}
They were adjusted to satisfactorily reproduce the average mass ${\overline M} = (M + 3 M^*)/4$ = 2008 MeV of
the $D$ mesons and the mass 2.087 MeV of $D_s$.


After integration in the flavor space, the two-body matrix elements  
containing contributions due to light,  strange and charm quarks are \cite{Stancu:2019qga}
\begin{equation}\label{twobody}
V_{ij} =
{\vec {\sigma}}_i\ \cdot {\vec {\sigma}}_j\, 
\left\{ \renewcommand{\arraystretch}{2}
\begin{array}{cl}
 V_{\pi} + \frac{1}{3} V^{uu}_{\eta} + \frac{1}{6} V^{uu}_{\eta_c},  &\hspace{0.3cm} \mbox{ $[2]_F, I = 1$} \\
 2 V_K - \frac{2}{3} V^{us}_{\eta}, ~~~ 2 V^{uc}_D- \frac{1}{2} V^{uc}_{\eta_c} & \hspace{0.3cm} \mbox{ $[2]_F, I = \frac{1}{2}$} \\
 2 V^{sc}_{D_s} - \frac{1}{2} V^{sc}_{\eta_c} & \hspace{0.3cm} \mbox{ $[2]_F, I = 0$} \\
 \frac{4}{3} V^{ss}_{\eta} + \frac{3}{2} V^{cc}_{\eta_c} & \hspace{0.3cm} \mbox{ $[2]_F, I = 0$} \\
-2 V^{sc}_{D_s} - \frac{1}{2} V^{sc}_{\eta_c} & \hspace{0.3cm} \mbox{ $[11]_F, I = 0$} \\
-2 V_K - \frac{2}{3} V^{us}_{\eta},~~~ - 2 V^{uc}_D - \frac{1}{2} V^{uc}_{\eta_c} &\hspace{0.3cm} \mbox{ $[11]_F, I = \frac{1}{2}$} \\
-3 V_{\pi} + \frac{1}{3} V^{uu}_{\eta} + \frac{1}{6} V^{uu}_{\eta_c},  &\hspace{0.3cm} \mbox{ $[11]_F, I = 0$}
\end{array} \right.
\end{equation}
\noindent
In Eqs. (\ref{twobody}) the pair  of quarks $ij$ is either in a symmetric [2]$_F$ or in an antisymmetric [11]$_F$ flavor state
and the isospin $I$ is defined by the quark content. The upper index of 
$V$ exhibits the flavor of the
two quarks interchanging a meson specified by the lower index.
In order to keep close to the notations of Ref.  \cite{Glozman:1995fu} the upper index of $\pi$ and 
$K$ is not indicated. Obviously, in every sum/difference of Eq. (\ref{twobody}) the upper index is the same
for all terms.

To calculate the matrix elements of the hyperfine interaction (\ref{VCHI}) between quarks
the first step is to decouple the flavor and spin parts of the wave function of partition $[f]_{FS}$
by using Clebsch-Gordan coefficients of the permutation group $S_4$ \cite{Stancu:1999qr}.
With the usual spin wave functions  and the flavor wave functions given in 
Appendix  \ref{flavor4},
one can reduce the calculation of four-body to two-body matrix elements.
Implementing the expressions (\ref{twobody}) 
one obtains
the matrix elements of the flavor-spin interaction (\ref{VCHI})
for four quark states in the flavor-spin space. The diagonal matrix elements are presented  in Table \ref{FOURQ}.

\begin{table}
\parbox{16cm}{\caption[matrix]{\label{FOURQ}
The hyperfine interaction $ V_{\chi}  $, Eq. (\ref{VCHI}), 
integrated in the flavor-spin space, for the quark subsystem $udsc$ with
$I$ = 0. 
${{V}^{q_a q_b}_{\gamma }}$ are defined in Eq.  (\ref{twobody})
where the upper index ${q_a q_b}$ indicates the flavor of the interacting quark pair.}} 
\begin{tabular}{c|c}
\hline
  State &   $  V_{\chi}  $ \\[0.5ex]
\hline
$| 1 \rangle = |{{\left[{31}\right]}_{O}\
{\left[{22}\right]}_{F}{\left[{22}\right]}_{S}{\left[{4}\right]}_{FS}}\rangle$
  & \hspace{1mm} 9 $ V_{\pi} -  V^{uu}_{\eta} - 2 V^{uu}_{\eta'} - \frac{1}{2} V^{uu}_{\eta_c}
 + 6 V_K + 6 V^{uc}_D + 6 V^{sc}_{D_s} + \frac{3}{2} V^{sc}_{\eta_c} - 2  V^{sc}_{\eta'} $ 
\\[1.2ex]
$| 2 \rangle = |{{\left[{31}\right]}_{O}\
{\left[{31}\right]}_{F}{\left[{31}\right]}_{S}{\left[{4}\right]}_{FS}}\rangle$
 & \hspace{1mm} 9 $V_{\pi} - V^{uu}_{\eta} - 2 V^{uu}_{\eta'} - \frac{1}{2} V^{uu}_{\eta_c} + 6 V_K
 + 6 V^{uc}_D + 2 V^{sc}_{D_s}  - \frac{1}{2} V^{sc}_{\eta_c}  + \frac{2}{3} V^{sc}_{\eta'}$ 
\\[1.2ex]
$| 3 \rangle =  |{{\left[{4}\right]}_{O}\
{\left[{211}\right]}_{F}{\left[{22}\right]}_{S}{\left[{31}\right]}_{FS}}\rangle$
  & \hspace{1mm}  $ \frac{14}{3} V_{\pi} - \frac{14}{27} V^{uu}_{\eta} - \frac{28}{27} V^{uu}_{\eta'}
- \frac{7}{27} V^{uu}_{\eta_c} +  \frac{14}{9} V_K + \frac{14}{27} V^{us}_{\eta} - \frac{14}{27} V^{us}_{\eta'}$
\\[0.9ex]
 & \hspace{1mm} + $\frac{46}{9}V^{uc}_D  + \frac{23}{18} V^{uc}_{\eta_c}  - \frac{46}{27} V^{uc}_{\eta'}
  +\frac{20}{9} V^{sc}_{D_s} + \frac{5}{9}  V^{sc}_{\eta_c} - \frac{20}{27} V^{sc}_{\eta'} $ \\[1.4ex]
$| 3' \rangle = |{{\left[{4}\right]}_{O}\
{\left[{211}\right]}^{'}_{F}{\left[{22}\right]}_{S}{\left[{31}\right]}_{FS}}\rangle$
  & \hspace{1mm}  $ \frac{13}{3} V_{\pi} - \frac{13}{27} V^{uu}_{\eta}  - \frac{13}{54} V^{uu}_{\eta_c} - \frac{26}{27} V^{uu}_{\eta'}  + \frac{20}{9}V^{uc}_D  + \frac{5}{9} V^{uc}_{\eta_c} - \frac{20}{27} V^{uc}_{\eta'}$
\\[0.9ex]
  & \hspace{1mm} + $\frac{52}{9} V_K + \frac{52}{27} V^{us}_{\eta} - \frac{52}{27} V^{us}_{\eta'}
 + \frac{10}{9} V^{sc}_{D_s} + \frac{5}{18}  V^{sc}_{\eta_c} - \frac{20}{54} V^{sc}_{\eta'} $ \\[1.4ex]
$| 4 \rangle = |{{\left[{4}\right]}_{O}\
{\left[{31}\right]}_{F}{\left[{22}\right]}_{S}{\left[{31}\right]}_{FS}}\rangle$
  & \hspace{1mm}  $ 6 V_{\pi} - \frac{2}{3} V^{uu}_{\eta} - \frac{4}{3} V^{uu}_{\eta'}  - \frac{1}{3} V^{uu}_{\eta_c}
 + \frac{2}{3}V^{uc}_D  + \frac{5}{6} V^{uc}_{\eta_c} - \frac{10}{9} V^{uc}_{\eta'}$
\\[0.9ex]
  & \hspace{1mm} + $\frac{2}{3} V_K + \frac{10}{9} V^{us}_{\eta} - \frac{10}{9} V^{us}_{\eta'}
 - \frac{4}{3} V^{sc}_{D_s} + \frac{1}{3}  V^{sc}_{\eta_c} - \frac{4}{9} V^{sc}_{\eta'} $ \\[1.4ex]
\hline
\end{tabular}
\end{table}

In the case of  $udsc\overline{c}$ pentaquarks there are also non-vanishing off-diagonal matrix elements.
These are
\begin{eqnarray}\label{off3andprime3}
{\langle 3 |  V_{\chi} | 3' \rangle} &=& \frac{\sqrt{2}}{9} 
( - 3 V_{\pi} + \frac{1}{3} V^{uu}_{\eta} +  \frac{1}{6} V^{uu}_{\eta_c} + \frac{2}{3} V^{uu}_{\eta'}  \nonumber \\
& + & 2 V_K + \frac{2}{3} V^{us}_{\eta} - \frac{2}{3} V^{us}_{\eta'} \nonumber \\
& + & 10 V^{uc}_D + \frac{5}{2} V^{uc}_{\eta_c} - \frac{2}{3} V^{uc}_{\eta'} \nonumber \\
& - & 10 V^{sc}_{D_s} - \frac{5}{2}  V^{sc}_{\eta_c} + \frac{2}{3} V^{sc}_{\eta'}),
\end{eqnarray}

\begin{eqnarray}\label{off3and4}
{\langle 3 |  V_{\chi} | 4 \rangle}    &=&  \frac{1}{2}
( - 6 V_{\pi} + \frac{2}{3} V^{uu}_{\eta} +  \frac{1}{3} V^{uu}_{\eta_c} + \frac{4}{3} V^{uu}_{\eta'}  \nonumber \\
& - & 8 V_K + V^{us}_{\eta} + \frac{4}{3} V^{us}_{\eta'} \nonumber \\
& - & 4 V^{uc}_D - 2 V^{uc}_{\eta_c} - \frac{4}{3} V^{uc}_{\eta'} \nonumber \\
& + & 4 V^{sc}_{D_s} - V^{sc}_{\eta_c} + \frac{4}{3} V^{sc}_{\eta'}),
\end{eqnarray}
and
\begin{eqnarray}\label{offprime3and4}
{\langle 3' |  V_{\chi} | 4 \rangle}       &=&  \sqrt{2}
( - 3 V_{\pi} + \frac{1}{3} V^{uu}_{\eta} +  \frac{1}{6} V^{uu}_{\eta_c} + \frac{2}{3} V^{uu}_{\eta'}  \nonumber \\
& + & 2 V_K + \frac{2}{3}V^{us}_{\eta} - \frac{2}{3} V^{us}_{\eta'} \nonumber \\
& - & 2 V^{uc}_D + \frac{1}{2} V^{uc}_{\eta_c} - \frac{2}{3} V^{uc}_{\eta'} \nonumber \\
& + & 2 V^{sc}_{D_s} - \frac{1}{2} V^{sc}_{\eta_c} + \frac{2}{3} V^{sc}_{\eta'}).
\end{eqnarray}
Note that the integration in the orbital space is not yet performed in the diagonal
and off-diagonal matrix elements presented above.


To reproduce the exact SU(4) limit   
one has to take $V_{\pi}$ = $V^{uu}_{\eta}$ 
= $V^{uu}_{\eta_c}$ = $V^{uc}_D$ = $V^{uc}_{\eta_c}$ = $V_K$ = $V^{sc}_{D_s}$ = $V^{sc}_{\eta_c}$ =  - ${C}_{\chi }$,
$V^{us}_{\eta}$ = - 3/4 ${C}_{\chi }$
and $V^{uu}_{\eta'}$ = $V^{uc}_{\eta'}$ = $V^{us}_{\eta'}$ = $V^{sc}_{\eta'}$ = 0. 
Then, in the exact SU(4) limit, the flavor-spin interaction 
takes the following form  \cite {Stancu:2019qga}
\begin{equation}\label{schematic}
{V}_{\chi }\ =\ -\ {C}_{\chi }\ \sum\limits_{ i\ <\ j}^{} {\lambda }_{
i}^{ F} \cdot {\lambda }_{ j}^{ F} \ {\vec {\sigma }}_{i} \cdot {\vec {\sigma}}_{j},
\end{equation}
with 
$C_{\chi}$
an equal strength constant for all pairs. 
Using Appendix \ref{Casimir},
one can check that the diagonal matrix elements   of Table \ref{FOURQ}
are - 27 $C_{\chi}$, - 21 $C_{\chi}$, - 15 $C_{\chi}$, - 15 $C_{\chi}$ and - 7 $C_{\chi}$ respectively. 
In the exact SU(4) limit the off-diagonal matrix elements of $ V_{\chi}$ vanish identically.
Thus the lowest state of Table \ref{FOURQ} is $|1 \rangle$ because it acquires the largest
attraction due to the FS interaction in the exact SU(4) limit. This
implies that the  lowest state 
has positive parity, conclusion  which sometimes still hold at broken symmetry, as for example for the 
$uudd \bar c$ pentaquarks \cite{Stancu:1998sm}.


\section{Orbital space}\label{coord}

The orbital wave functions are defined in terms of 
four internal Jacobi coordinates for pentaquarks chosen as
\begin{eqnarray}\label{coordin}
\begin{array}{c}\vec{x}\ =\ {\vec{r}}_{1}\ -\ {\vec{r}}_{2}\ ,\, \\ 
\vec{y}\ =\
{\left({{\vec{r}}_{1}\ +\ {\vec{r}}_{2}\ -\ 2{\vec{r}}_{3}}\right)/\sqrt
{3}} ,\, \\
\vec{z}\ =\ {\left({{\vec{r}}_{1}\ +\ {\vec{r}}_{2}\ +\ {\vec{r}}_{3}\ -\
3{\vec{r}}_{4}}\right)/\sqrt {6}} ,\, \\  
\vec{t}\ =\
{\left({{\vec{r}}_{1}\
+\ {\vec{r}}_{2}\ +\ {\vec{r}}_{3}+\ {\vec{r}}_{4}-\
4{\vec{r}}_{5}}\right)/\sqrt {10}} ,
\end{array}
\end{eqnarray}
where 1,2,3 and 4 are the quarks and 5 the antiquark so that
$t$ gives the distance between the antiquark and the center of mass coordinate 
of the four-quark subsystem.

For the lowest positive parity states having $\ell$ = 1,
there are two ways to introduce orbital excitations \cite{Stancu:2019qga}.
One is to excite the four-quark subsystem, the other is to include the angular 
momentum in the relative motion between the four-quark subsystem and the antiquark.
Both imply translational invariant states (no center of mass motion).

\subsection{Excited four-quark subsystem, $P$ = + 1}{Liu:2019tjn

In this case one has to 
express the orbital wave functions of the four-quark subsystem of structure $s^3p$
in terms of the internal coordinates 
$\vec{x},  \vec{y},  \vec{z}$ 
for the specific permutation symmetry
${\left[{31}\right]}_{O}$.
The method of
constructing translationally invariant states of definite permutation symmetry
containing a unit of angular momentum was first given in Ref. \cite{Stancu:1998sm}
and recently revised in  Ref. \cite{Stancu:2019qga}.
The 
three independent states  denoted below by $\psi_i$, which define the 
basis vectors   of the irreducible representation 
${\left[{31}\right]}_O$   in 
terms of shell model states 
$ \left\langle{\vec{r}\left|{n \ell m}\right.}\right\rangle$ where $n$ = 0, $\ell$ = 1, 
are
\begin{eqnarray}
{\psi }_{1} = 
\left\langle{\vec{x}\left|{000}\right.}\right
\rangle\left\langle{\vec{y}\left|{000}\right.}\right\rangle\left
\langle{\vec{z}\left|{010}\right.}\right\rangle
\end{eqnarray}
\begin{eqnarray}
{\psi }_{2} = 
\renewcommand{\arraystretch}{0.5}
\left\langle{\vec{x}\left|{000}\right.}\right
\rangle\left\langle{\vec{y}\left|{010}\right.}\right\rangle\left
\langle{\vec{z}\left|{000}\right.}\right\rangle
\end{eqnarray}
\begin{eqnarray}
{\psi }_{3} = 
\left\langle{\vec{x}\left|{010}\right.}\right
\rangle\left\langle{\vec{y}\left|{000}\right.}\right\rangle\left
\langle{\vec{z}\left|{000}\right.}\right\rangle
\end{eqnarray}


\noindent 
In this picture there is no
excitation in the relative motion between the cluster of four quarks and the 
antiquark defined by the coordinate $\vec{t}$. Then the  
pentaquark orbital wave functions $\psi_i^5$ are  obtained by multiplying
each $\psi_i$ from above by the wave function
$\left\langle{\vec{t}\left|{000}\right.}\right\rangle$ which describes the
relative motion between the four-quark subsystem and the antiquark $\overline{c}$.
Assuming an exponential behavior we introduce two variational parameters, 
$a$ for the internal motion of the four-quark subsystem  and
$b$ for the relative motion between the subsystem $qqqc$ and $\overline{c}$. We explicitly have
\begin{equation}\label{psi1}
{\psi }_{1}^{5}\ =\ N\ \exp\ \left[{-\ {\frac{a}{2}}\ \left({{x}^{2}\ +\
{y}^{2}\ +\ {z}^{2}}\right)\ -\ {\frac{b}{2}}\ {t}^{2}}\right]\ z\ {Y}_{10}\
\left({\hat{z}}\right)
\end{equation}
\begin{equation}\label{psi2}
{\psi }_{2}^{5}\ =\ N\ \exp\ \left[{-\ {\frac{a}{2}}\ \left({{x}^{2}\ +\
{y}^{2}\ +\ {z}^{2}}\right)\ -\ {\frac{b}{2}}\ {t}^{2}}\right]\ y\ {Y}_{10}\
\left({\hat{y}}\right)
\end{equation}
\begin{equation}\label{psi3}
{\psi }_{3}^{5}\ =\ N\ \exp\ \left[{-\ {\frac{a}{2}}\ \left({{x}^{2}\ +\
{y}^{2}\ +\ {z}^{2}}\right)\ -\ {\frac{b}{2}}\ {t}^{2}}\right]\ x\ {Y}_{10}\
\left({\hat{x}}\right)
\end{equation}
where
\begin{equation}
N\ =\ {\frac{{2}^{3/2}{a}^{11/4}{b}^{3/4}}{{3}^{1/2}{\pi }^{5/2}}}
\end{equation}


\subsection{Excitation between the four-quark subsystem and the antiquark, $P$ = +1}

The authors of Ref. \cite{Yuan:2012wz} have studied the $qqqc \bar c$ and the $qqsc \bar c$ pentaquarks, 
in three different
models, including  the FS model. 
The orbital wave function of the four-quark subsystem has symmetry $[4]_O$ for both 
parities. 
Although the radial wave function was not specified, one can infer that the positive parity states of Ref.  \cite{Yuan:2012wz} 
were obtained by including a unit of orbital angular momentum in the relative motion  
between the four-quark subsystem and the antiquark. The states remains translationally invariant.
In this case the orbital wave function takes the form
\begin{equation}\label{psi4}
{\psi }_{4}^{5}\ =\ N_4 \exp\ \left[{-\ {\frac{a}{2}}\ \left({{x}^{2}\ +\
{y}^{2}\ +\ {z}^{2}}\right)\ -\ {\frac{b}{2}}\ {t}^{2}}\right]\ t\ {Y}_{10}\
\left({\hat{t}}\right),
\end{equation}
where
\begin{equation}
N_4 =\ \frac{8^{1/2} a^{9/4} b^{5/4}}{3^{1/2} \pi^{5/2}}. 
\end{equation}

\subsection{Negative parity states, $P$ = - 1}

We also need the orbital wave function of 
the lowest negative parity state described by the
$s^4$ configuration of symmetry $[4]_O$  which is
\begin{equation}\label{phi}
\phi_0 = \ N_0\ \exp\ \left[{-\ {\frac{a}{2}}\ \left({{x}^{2}\ +\
{y}^{2}\ +\ {z}^{2}}\right)\ -\ {\frac{b}{2}}\ {t}^{2}}\right],
\end{equation}
with
\begin{equation}\label{phinorm}
N_0\ =\ ({\frac{a}{\pi}})^{9/4} (\frac{b}{\pi})^{3/4}.
\end{equation}


\section{Kinetic energy}\label{ke}

The kinetic energy  $T$  of the Hamiltonian (\ref{ham})
can be calculated analytically. Below we present the expression of its 
expectation value for the three cases introduced above.

\vspace{1cm}
{\it Case A.}
In this case the  expectation value of the kinetic energy is defined by the average over the three wave functions
defined by Eqs. (\ref{psi1})-(\ref{psi3}). One obtains
\begin{eqnarray}\label{TA}
\begin{array}{lcl}\left\langle{T}\right\rangle\ &=&\ {\frac{1}{3}}\
\left[{\left\langle{{\psi }_{1}^{5}\left|{T}\right|{\psi\
}_{1}^{5}}\right\rangle\ +\ \left\langle{{\psi }_{2}^{5}\left|{T}\right|{\psi
}_{2}^{5}}\right\rangle\ +\ \left\langle{{\psi }_{3}^{5}\left|{T}\right|{\psi
}_{3}^{5}}\right\rangle}\right]\\
\\ &=&\ {\hbar }^{2}\ \left({{\frac{11}{2{\mu }_{1}}}\ a\ +\ {\frac{3}{{2\mu
}_{2}}}\ b}\right),
\end{array}
\end{eqnarray}
with
\begin{eqnarray}\label{reduced}
{\frac{4}{{\mu }_{1}}}\
 =  \ \frac{2}{{m}_{q}}\  + \ \frac{1}{{m}_{s}}\ + \ \frac{1}{{m}_{Q}},\ 
\end{eqnarray}
which is the generalization  of Eq. (22) of Ref. \cite{Stancu:2019qga} to include strange quarks
and  
\begin{equation}\label{redmass}
{\frac{5}{{\mu }_{2}}}\ =\ {\frac{1}{{\mu }_{1}}}\ +\ {\frac{4}{{m}_{Q}}},
\end{equation}
where $q = u, d$ and $Q = c$.
Here, we have  $ m_q $ = 340 MeV,  $ m_s $ = 440 MeV and $m_c$ = 1350 MeV, 
as defined by Eq. (\ref{quarkmass}). 
Taking $m_u$ = $m_d$  = $m_s$ =  $m_Q$ = $m$ and setting $a$ = $b$, one can recover the identical particle limit
$\left\langle{T}\right\rangle\ =\ {\frac{7}{2}}\ \hbar \omega$ with
$\hbar \omega  \ =\ 2\ a{\hbar }^{2}/m$. 

\vspace{1cm}
{\it Case B.}
In this case there is only one orbital wave function because we deal with the symmetric 
state $[4]_O$. The orbital excitation is located in the relative motion of the four-quark system 
and the antiquark. One obtains
\begin{eqnarray}\label{TB}
\left\langle{T}\right\rangle\ = \ {\hbar }^{2}\ \left({{\frac{9}{2{\mu }_{1}}}\ a\ +\ {\frac{5}{{2\mu
}_{2}}}\ b}\right),
\end{eqnarray}
where $\mu_{1}$ and $\mu_{2}$ are the same as above. Again one can recover the identical particle limit
when $a$ = $b$ but the contributions of the two terms are  different because the 
coefficients 11/2 and 3/2 now become 9/2 and 5/2 respectively,
which is natural because the unit of orbital excitation is no more located in the four quark subsystem but
in the relative motion between the four-quark subsystem and $\bar c$.

\vspace{1cm}
{\it Case C.}
One deals with the symmetric state $[4]_O$ and no orbital excitation. The only orbital state has negative parity 
and  Eq. (\ref{phi}) gives
\begin{eqnarray}\label{TC}
\left\langle{T}\right\rangle\ = \ {\hbar }^{2}\ \left({{\frac{9}{2{\mu }_{1}}}\ a\ +\ {\frac{3}{{2\mu
}_{2}}}\ b}\right),
\end{eqnarray}
with $\mu_{1}$ and $\mu_{2}$  as above.

\section{Confinement}\label{linearcon}

By integrating in the color space, the expectation value of the confinement interaction (\ref{conf}) 
has the same form as that of the $uudc \bar c $ system  \cite{Stancu:2019qga}
\begin{equation}\label{confin}
\left\langle{{V}_{conf}}\right\rangle\ =\ {\frac{C}{2}}\ \left({6\
\left\langle{{r}_{12}}\right\rangle\ +\ 4\
\left\langle{{r}_{45}}\right\rangle}\right)
\end{equation}
where $\langle{{r}_{ij}}\rangle$ is the interquark distance and
the coefficients 6 and 4 account for the number of quark-quark and
quark-antiquark pairs, respectively,  for all cases $A$, $B$ and $C$,
but with different expressions for $\langle{{r}_{ij}}\rangle$ in each case.

\vspace{1cm}
{\it Case A.}
Here one has
\begin{equation}\label{r12}
\left\langle{{r}_{ij}}\right\rangle\ =\ {\frac{1}{3}}\
\left[{\left\langle{{\psi }_{1}^{5}\left|{{r}_{ij}}\right|{\psi
}_{1}^{5}}\right\rangle\ +\ \left\langle{{\psi
}_{2}^{5}\left|{{r}_{ij}}\right|{\psi }_{2}^{5}}\right\rangle\ +\
\left\langle{{\psi }_{3}^{5}\left|{{r}_{ij}}\right|{\psi
}_{3}^{5}}\right\rangle}\right],
\end{equation}
where $i,j$ = 1,2,3,4,5 ($ i \neq j$).
An analytic evaluation gives
\begin{equation}
\left\langle{{r}_{12}}\right\rangle\ =\ {\frac{20}{9}}\ \sqrt {{\frac{1}{\pi
 a}}},
\end{equation}
and 
\begin{equation}
\left\langle{{r}_{45}}\right\rangle\ =\ {\frac{1}{3\sqrt {2\pi }}}\
\left[{2\sqrt {{\frac{3}{a}}\ +\ {\frac{5}{b}}}\ +\ \sqrt {5b}\
\left({{\frac{1}{2a}}\ +\ {\frac{1}{b}}}\right)}\right].
\end{equation}

\vspace{1cm}
{\it Case B.}
The expectation value of the confinement interaction is given by Eq. (\ref{confin})
with
\begin{equation}
\left\langle{{r}_{12}}\right\rangle\ =\ \sqrt {{\frac{4}{\pi
 a}}},
\end{equation}
and
\begin{equation}
\left\langle{{r}_{45}}\right\rangle\ =\ {\frac{2}{3} \sqrt{\frac{2\ b}{5\ \pi }}}\ 
\left({{\frac{3}{4a}}\ +\ {\frac{5}{b}}}\right).
\end{equation}

\vspace{1cm}
{\it Case C.} 
In this case  the four quarks are in the
$s^4$ configuration described by the states $| 3  \rangle$, $| 3'  \rangle$ or $| 4  \rangle$
and there is no orbital excitation at all.
The expectation value of the confinement interaction is given by Eq. (\ref{confin})
as well, with 
\begin{equation}\label{r12ground}
\left\langle{{r}_{12}}\right\rangle\ = \sqrt {{\frac{4}{\pi a}}},
\end{equation}
and 
\begin{equation}\label{r45ground}
\left\langle{{r}_{45}}\right\rangle\ =\ {\frac{1}{\sqrt {2\pi }}}\
\sqrt{{\frac{3}{a}}\ +\ {\frac{5}{b}}}.
\end{equation}


\section{Flavor-spin interaction}\label{FSint}

In order to integrate the expressions  of  Table \ref{FOURQ} and Eqs. (\ref{off3andprime3})-(\ref{offprime3and4})
in the orbital space one has to decouple 
the orbital part of the wave function $[f]_O$ from the part containing the other degrees of
freedom by using Clebsch-Gordan coefficients of the permutation group $S_4$ \cite{Stancu:1999qr}.
The next step is to reduce 
the matrix elements of the hyperfine interaction $V^{\chi}$ of Eq. (\ref{VCHI}) 
of the four quark system to matrix elements of two quarks.
Table \ref{FOURQ} gives the diagonal matrix elements and 
Eqs. (\ref{off3andprime3})-(\ref{offprime3and4}) the off-diagonal ones. 
As there are 6 pairs, the contribution of one pair is
one sixth of the above expressions.

For states of type $A$  with one unit of orbital excitation  
the result is a linear combination of  orbital two-body matrix elements of type
$\left\langle{ss\left|{{V}^{q_a q_b}_{\gamma }}\right|ss}\right\rangle\ ,\
\left\langle{sp\left|{{V}^{q_a q_b}_{\gamma }}\right|sp}\right\rangle$ and
$\left\langle{sp\left|{{V}^{q_a q_b}_{\gamma }}\right|ps}\right\rangle$.
For states of type $B$ or $C$ there are  two-body matrix elements between single particle $s$-states, namely
$\left\langle{ss\left|{{V}^{q_a q_b}_{\gamma }}\right|ss}\right\rangle$.
In every term  $q_a q_b$
is a pair of quarks from Eq. (\ref{twobody}).

%
\section{Results and discussion}\label{numerical}

We have looked for variational solutions of the Hamiltonian of Sec. \ref{Hamiltonian}
using the orbital part of the wave functions as
described in Sec. \ref{coord}, which contain the parameters $a$ and $b$.
The wave functions are  the product of the four quarks subsystem states 
of flavor-spin structure defined in Table \ref{FOURQ}  and the
charm antiquark wave function denoted by $|\bar c \rangle$. 
The total angular momentum is  $\vec{J}\ =\ \vec{L}\ +\ \vec{S}\ +\
{\vec{s}}_{Q}$, with $\vec{L}$ and $\vec{S}$ the angular momentum and spin of the
four-quark cluster and $\vec{s}_Q$ the spin of the heavy antiquark.

We have neglected the contribution of  $V^{uu}_{\eta_c}$, $V^{uc}_{\eta_c}$ and $V^{sc}_{\eta_c}$  because little
$u \bar u$, $d \bar d$ and $s \bar s$ are expected in ${\eta_c}$.   We have also neglected 
$V^{uc}_{\eta'}$ and  $V^{sc}_{\eta'}$ assuming a little $c \bar c$ component in $\eta'$. 
Thus, in the expressions of Table \ref{FOURQ} we took 
\begin{equation}
V^{uu}_{\eta_c} = V^{uc}_{\eta_c} = V^{uc}_{\eta'} =  V^{sc}_{\eta_c} =  V^{sc}_{\eta'} =  0.
\end{equation}

For Case $A$ the numerical results are presented in Table \ref{resultsA}.  
The eigenvalues of  $|1 \rangle |\bar c \rangle $
and    $|2 \rangle |\bar c \rangle $  states are degenerate for the allowed values of $J$ in each case.  
For  $|2 \rangle |\bar c \rangle $ the states with $J^P =  {1/2}^+$ and  ${3/2}^+$
have multiplicity 2. 
The optimal values found for 
the parameters $a$ and $b$ are the same for both states. We found that the ratio of the matrix elements
of the $K$- and
$\pi$-meson exchange is about 0.74, close to the quark mass ratio $m_{u,d}/m_s$ and the matrix elements 
of the $K$- and $D$-meson exchange is about 0.34 close to the ratio $m_s/m_c$. 


\begin{table}
\caption{\label{resultsA}Lowest positive  parity  $udsc \bar c$ pentaquarks
of quantum numbers $S$ and  $J^P$ and symmetry structure $|1 \rangle$ and $|2 \rangle$
defined in Table \ref{FOURQ}.
Column 1 gives the state, column 2 the spin, column 3  the parity and total angular momentum, 
column 4 the
optimal variational parameters associated to the wave functions defined in Sec. \ref{coord},
and column 5 the calculated mass.}
\begin{tabular}{cccccccc}
\hline
State & \hspace{3mm}$S$ & $J^P$ & \multicolumn{2}{c}{Variational parameters} &
Mass   \\
&\hspace{6mm} & \hspace{6mm} & a (fm$^{-2}$) & b (fm$^{-2}$) & (GeV) \\
\hline 
$ | 1 \rangle ~|\overline c \rangle$ & \hspace{1mm} $\frac{1}{2}$ & \hspace{2mm}  $\frac{1}{2}^+$, $\frac{3}{2}^+$ 
& \hspace{6mm} 1.798 & 1.053 & 4442  \\[1.1ex]
$ | 2 \rangle ~|\overline c\rangle$  & \hspace{1mm} $\frac{1}{2}$ & \hspace{2mm}  $\frac{1}{2}^+$,\hspace{1mm} $\frac{3}{2}^+$,\hspace{1mm} $\frac{5}{2}^+$ & \hspace{6mm} 1.798 & 1.053 & 4495  \\[1.1ex]
\hline
\end{tabular}
\end{table}


\begin{table}
\parbox{16cm}{\caption[matrix]{\label{diag3positive}
The mass and the mixing coefficients of states of positive parity
$|3 \rangle |\bar c \rangle $, $|3' \rangle  |\bar c \rangle $ and  $|4 \rangle |\bar c \rangle  $ defined in Table \ref{FOURQ}
with $L$ = 1, $S$ = 0, $J^P$ = $1/2^{+}$, $3/2^{+}$ obtained from the orbital
wave function of Case $B$ with $a$ = 1.798 $fm^{-2}$ and $b$ = 1.053 $fm^{-2}$.}} 
\begin{tabular}{c|c}
\hline
  Mass (MeV)   &  \hspace{6mm} $|3 \rangle|\bar c \rangle $ \hspace{1.2cm} $|3' \rangle |\bar c\rangle $ \hspace{1.2cm} $|4 \rangle |\bar c \rangle$  \\[0.5ex]
\hline
4493           &  \hspace{6mm} 0.748  \hspace{6mm}  0.324   \hspace{6mm}  -0.579 \\[0.5ex]
4614           &  \hspace{6mm} 0.326  \hspace{6mm} -0.939   \hspace{6mm}  -0.104 \\[0.5ex]
5075           & \hspace{6mm} -0.578  \hspace{6mm} -0.111   \hspace{6mm}  -0.808 \\[1.2ex]
\hline
\end{tabular}
\end{table} 

For Case $B$ the masses and the mixing coefficients of the  ${1/2}^+$ and  ${3/2}^+$ states,
obtained from the combination of the basis vectors  $|3 \rangle |\bar c \rangle $,  $|3' \rangle |\bar c \rangle $
and $|4 \rangle |\bar c \rangle $ are presented in Table \ref{diag3positive}.
The optimal variational parameters are the same as in Table \ref{resultsA}.
The mixing coefficients turn to be all large for the lowest state of 4493 MeV. The next state
at 4614 MeV is dominantly a $|3' \rangle |\bar c \rangle $ state and the last eigenstate  at 5075 is mostly
a combination of   $|3 \rangle |\bar c \rangle $ and $|4 \rangle |\bar c \rangle $
due to the large off-diagonal matrix element  (\ref{offprime3and4}) where the dominant
$\pi$- and $K$-meson exchanges contribute with the same sign.

The  Case $C$ corresponding to negative parity ${1/2}^-$ state
is shown in Table  \ref{ground3negative}.
The mixing coefficients are the same as those of Table \ref{diag3positive}, because they result from the
diagonalization of a hyperfine interaction  identical to that of Case $B$.
The difference between these cases appears
only in the kinetic and the confinement matrices, which are diagonal. Hence, in Case $C$  
the masses 
can be obtained from those  of Table \ref{diag3positive}
by lowering each of them by 89 MeV which is precisely the difference in the 
kinetic energy plus the confinement energy between Case $B$ and Case $C$. 
The largest mixing is between the states $|3 \rangle |\bar c \rangle $ and $|4 \rangle |\bar c \rangle $.
The diagonal matrix element of the Hamiltonian $\langle 3  \bar c| H |3 \bar c \rangle $  
is lowered from 4612 MeV to 4404 MeV and the value of  $\langle 4  \bar c| H |4 \bar c \rangle $
is increased from 4786 MeV to 4986 MeV.

Looking at Tables  \ref{resultsA},~\ref{diag3positive} and  \ref{ground3negative} one can see that 
the lowest mass is 4404 MeV. Thus the lowest pentaquark  $udsc \overline{c}$ has quantum
numbers $J^P$ = ${1/2}^-$, in contrast to the lowest pentaquark  $uudc \overline{c}$
for which it was found  $J^P$ = ${1/2}^+$ in Ref. \cite{Stancu:2019qga}.

The mixing of states $|3 \rangle |\bar c \rangle $,  $|3' \rangle |\bar c \rangle $
and $|4 \rangle |\bar c \rangle $ has been first discussed in Ref. \cite{Yuan:2012wz}
with the corresponding notation $|3 \rangle \rightarrow |1 \rangle$,    $|3' \rangle \rightarrow |1' \rangle$   
and $|4 \rangle \rightarrow |2 \rangle$ where the quark model of Ref. \cite{Glozman:1995xy}
with a harmonic oscillator confinement and a simplified  hyperfine interaction have been used. 
The mixing was introduced  for $J^P$ = ${1/2}^-$ only, case $C$. There the  $J^P$ = ${1/2}^-$ state appears
at 4084 MeV and the  $J^P$ = ${1/2}^+$ state at 4291 MeV, {\it i. e.} about 200 MeV
above the lowest negative parity state. Thus the lowest  $J^P$ = ${1/2}^-$ state of Ref.  \cite{Yuan:2012wz}
is about 300 MeV lower than in the present case.

The  $J^P$ = ${1/2}^-$ states found in this study are located within the energy range of the
 $J^P$ = ${1/2}^-$ resonances predicted in Ref. \cite{Xiao:2019gjd}. There only $s$-wave meson-baryon interactions were
considered so that only negative parity states were discussed. Their coupling to the $J/\psi \Lambda$ channel 
was found to be small, but large enough to provide convenient production rates.
The masses of hidden charm strange pentaquarks with  $J^P$ = ${1/2}^-$ found in 
Ref. \cite{Wang:2019nvm} within a chiral effective field theory are located as well in the energy
range predicted in the present work.
A similar mass range was found in Ref. \cite{Ferretti:2020ewe} in a hadrocharmonium picture, 
with the difference that the lowest state has positive parity.

\begin{table}
\parbox{16cm}{\caption[matrix]{\label{ground3negative}
The mass and the mixing coefficients of states of negative parity, Case $C$,  diagonalized 
in the basis 
$|3 \rangle$, $|3' \rangle$ and  $|4 \rangle$ defined in Table \ref{FOURQ}
with $L$ = 0, $S$ = 0, $J^P$ = $1/2^{-}$. The variational parameters of  the orbital
wave function are  $a$ = 1.798 $fm^{-2}$ and $b$ = 1.053 $fm^{-2}$.}} 
\begin{tabular}{c|c}
\hline
  Mass (MeV)   &  \hspace{6mm} $|3 \rangle$  \hspace{1.2cm} $|3' \rangle$    \hspace{1.2cm} $|4 \rangle$  \\[0.5ex]
\hline
4404           &  \hspace{6mm} 0.748  \hspace{6mm}  0.324   \hspace{6mm}  -0.579 \\[0.5ex]
4525           &  \hspace{6mm} 0.326  \hspace{6mm} -0.939   \hspace{6mm}  -0.104 \\[0.5ex]
4986           & \hspace{6mm} -0.578  \hspace{6mm} -0.111   \hspace{6mm}  -0.808 \\[1.2ex]
\hline
\end{tabular}
\end{table} 


\section{Conclusions}

We have calculated a few of the lowest  masses  of the hidden charm strange 
pentaquarks $udsc \bar c$,  in  the SU(4) version of the flavor-spin model
introduced in Ref. \cite{Stancu:2019qga} where it was applied to $uudc \bar c$ 
pentaquarks.
The model provides an isospin dependence and an internal structure of pentaquarks. 
For positive parity the angular 
momentum can be located in the internal motion of the four-quark subsystem, Case $A$,   
or in the relative motion between the four-quark subsystem 
and the antiquark, Case $B$. 

According to the discussion 
presented in Ref.  \cite{Stancu:2019qga} at exact SU(4) symmetry the lowest 
positive pentaquark state has positive parity when the orbital excitation is located in the
internal motion of the four-quark subsystem. For broken SU(4) such a result remained 
valid for the $uudc \overline{c}$ pentaquark.
In the present analysis it was found that the lowest state of the $udsc \bar c$ pentaquark   has negative parity. 
This is due to the breaking of SU(4)-flavor symmetry which, coupling  states of different 
flavor symmetry $[f]_F$, lowers considerably the negative parity state and not so much the positive
parity ones. As a consequence, the negative parity state  $J^P$ = $1/2^{-}$, without any orbital excitation,  Case $C$,
was found to have the lowest mass of 4404 MeV, followed by the lowest positive 
parity states   $J^P$ = $1/2^{+}$ or $3/2^{+}$   with a mass of 4442 MeV.

There is an important difference between $udsc \overline{c}$ and   $uudc \overline{c}$ pentaquarks
due to the presence of the quark $s$.
The $udsc \overline{c}$ pentaquark
has two Weyl tableaux 
associated to the irreducible representation $[211]$ of the four-quark subsystem at $I$ = 0, as shown in Appendix \ref{flavor4}. 
Due to the Pauli principle the
$uudc \overline{c}$ pentaquark has only one Weyl tableau associated to the irreducible
representation $[211]$. Accordingly, in the  $udsc \overline{c}$ pentaquark there are 
three states which can couple due to the SU(4) breaking, the $|3 \rangle$,  $|3' \rangle$ and $| 4 \rangle$,
as shown in the present study. 
As mentioned above, this coupling brings the lowest $J^P$ = $1/2^{-}$ state below 
the lowest positive 
parity states   $J^P$ = $1/2^{+}$ or $3/2^{+}$.

In the  $uudc \overline{c}$ pentaquark, 
there are only two flavor states which, in principle, can couple due to
the breaking of SU(4). They are of type $|3 \rangle$ and $| 4 \rangle$ with  appropriate Weyl tableaux.
We found out that the coupling between 
the states of symmetry  
$| 3 \rangle =  |{{\left[{4}\right]}_{O}\
{\left[{211}\right]}_{F}{\left[{22}\right]}_{S}{\left[{31}\right]}_{FS}}\rangle$
and 
$| 4 \rangle = |{{\left[{4}\right]}_{O}\
{\left[{31}\right]}_{F}{\left[{22}\right]}_{S}{\left[{31}\right]}_{FS}}\rangle$
vanish identically for the $uudc \overline{c}$ pentaquark.
Therefore the lowest state in  the $uudc \overline{c}$ pentaquark
has positive parity, as shown in Ref.  \cite{Stancu:2019qga}. This conclusion is
at variance with the result of Ref. \cite{Yuan:2012wz} where  $| 3 \rangle$
and $| 4 \rangle$ mix together. A possible reason of the discrepancy is that the three flavor states 
of symmetry $[31]$, as defined by  Eqs. (A.9)-(A.11) of Ref. \cite{Yuan:2012wz} 
do not  form a proper Young Yamanouchi 
basis for the irreducible representation $[31]$ of the permutation group S$_4$.

We recall that the parity sequence of the  $uudc \overline{c}$ pentaquark 
studied in the hadrocharmonium model \cite{Eides:2019tgv} was similar to ours \cite{Stancu:2019qga},
namely that the lowest pentaquark state has $J^P$ = $1/2^+$ quantum numbers. 
In the hadrocharmonium description of Ref. \cite{Ferretti:2020ewe} 
the lowest state of the $udsc \overline{c}$ pentaquark 
has positive parity, contrary to the present result.

Therefore, in the flavor-spin model
the presence of the strange quark brings more richness to the flavor structure
and changes the parity order of the lowest two state in the $udsc \overline{c}$ pentaquark relative
to the $uudc \overline{c}$ pentaquark. 

The $J^P$ quantum numbers of the 2019 LHCb resonances are not yet known.
Likewise, for possible future observations the spin and parity will be essential
to discriminate  between the existing interpretations of pentaquarks, or
inspire new developments.


\appendix


\section{Exact SU(4) limit}\label{Casimir}

The exact SU(4) limit is useful in checking the integration in the flavor space,
made in Table  \ref{FOURQ}.  
In this limit every expectation value of  Table \ref{FOURQ}  reduces to the expectation value of Eq.  (\ref{schematic})
and one can use the following
formula \cite{Ortiz-Pacheco:2018ccl}
\begin{equation}\label{lambdasigma}
\langle ~\sum_{i<j} \lambda^F_i \cdot \lambda^F_j \vec{\sigma}_i \cdot \vec{\sigma}_j~\rangle = 4
C_{2}^{SU(2n)} - 2 C_{2}^{SU(n)} -
\frac{4}{k} C_{2}^{SU(2)} - k \frac{3(n^2 - 1)}{n}
\end{equation}
where $n$ is the number flavors and  $k$ the number of quarks, here $n$ = 4 and $k$ = 4.  $C_{2}^{SU(n)}$ is the
Casimir operator eigenvalues of $SU(n)$ which can be derived from the
expression  \cite{Stancu:1997dq} :
\begin{eqnarray}\label{casimiroperator}
C_{2}^{SU(n)} = \frac{1}{2} [f_1'(f_1'+n-1) + f_2'(f_2'+n-3) + f_3'(f_3'+n-5)
 \nonumber \\
+f_4'(f_4'+n-7) + ... + f_{n-1}'(f_{n-1}'-n+3) ] - \frac{1}{2n}
(\sum_{i=1}^{n-1} f_i')^2
\label{casimir}
\end{eqnarray}
where $f_i'= f_i-f_n$, for an irreducible representation given by the
partition $[f_1,f_2,...,f_n]$.
Eq. (\ref{lambdasigma}) has been previously used for $n$ = 3 and $k$ = 6 in Ref.  \cite{Stancu:1997dq}. 


\section{The baryons}\label{baryons}

The masses of ground state baryons relevant to the  study of $u d s c \bar c$ 
pentaquarks with isospin $I$ = 0 were estimated variationally by
using a radial wave function of the form 
$\phi \propto  exp[-\frac{a}{2}(x^2 + y^2)]$
containing the variational parameter  $a$
and the coordinates $x$ and $y$ defined by Eq. (\ref{coordin}).
The results are indicated in Table \ref{baryon} together with the experimental
masses.
We took $V^{uc}_{\eta_c} =   V^{uc}_{\eta'} =  V^{sc}_{\eta_c} =  V^{sc}_{\eta'} = 0$. 
The resulting charmed baryon masses  are about 100 MeV lower than
the experimental values. By increasing the charmed 
quark mass from $m_c$ = 1.35 GeV to $m_c$ = 1.45 GeV the agreement
with the experiment would be much better. 
However, we prefer to use the same parametres as in Ref. \cite{Stancu:2019qga}
in order to make a comparison with the $uudc \bar{c}$ pentaquarks.

\begin{table*}
\caption{\label{baryon} Masses of ground state baryons with the
flavor-spin interaction of Sec. \ref{Hamiltonian}.
Column 1 gives the baryon, column 2 the isospin, column 3 the spin and parity 
column 4 the calculated mass, column 5 the variational parameter 
and the last column the experimental mass.}
\begin{tabular}{ccccccc}
\hline
Baryon & \hspace{2mm}$I$ & $J^P$ & {Calc. Mass (GeV)}     & \hspace{1mm} a(fm$^{-2}$) & \hspace{0mm} Exp.mass (GeV)\\
\hline
$ \Lambda $ & \hspace{1mm} $ 0 $ & \hspace{2mm}  $\frac{1}{2}^+$ & \hspace{5mm} 1.165   & \hspace{2mm}  2.484 
&   1.116  \\[1.1ex]
$ \Lambda_c $ & \hspace{1mm} $ 0 $ & \hspace{2mm}  $\frac{1}{2}^+$ & \hspace{5mm} 2.180   & \hspace{2mm}  2.055 
&   2.283  \\[1.1ex]
$ \Xi_c     $ & \hspace{1mm} $ 0 $ & \hspace{2mm}  $\frac{1}{2}^+$ & \hspace{5mm} 2.304   & \hspace{2mm}  1.797 
&   2.469  \\[1.1ex] 
\hline
\end{tabular}
\end{table*}

\section{The flavor wave functions}\label{flavor4}

The four quark flavor states of content $udsc$ 
defining the basis vectors of the irreducible representations   $[{31}]_F$,  $[{22}]_F$, $[{211}]_F$ and $[{1111}]_F$
have been given in Ref. \cite{Yuan:2012wz} for I = 0. We have checked them with 
the method of Ref. \cite{Stancu:1991rc}. 
In Ref. \cite{Yuan:2012wz} the flavor states were defined in the 
Young-Yamanouchi basis. The order of particles is always 1234 in every term. 

In Table \ref{flavorstates}, except for $[{1111}]_F$, not needed here,  we give the correspondence 
between the Young-Yamanouchi basis and the
notation of Ref. \cite{Yuan:2012wz} for each Yamanouchi symbol
which is a compact notation for a Young tableau. For a tableau with $n$ particles it
is defined by $Y = (r_n,r_{n-1},...,r_1)$ where $r_i$ represents the row of the particle $i$.
The Weyl tableaux are indicated for each irreducible representation.

Here we write the
flavor states in terms of products of  symmetric 
$\phi_{[2]}(q_a q_b) = (q_a q_b + q_b q_a)/\sqrt{2} $ or  antisymmetric  
$\phi_{[11]}(q_a q_b) = (q_a q_b - q_b q_a)/\sqrt{2} $ 
quark pair states for the pairs 12 and 34.  This allows a straightforward 
calculation of the flavor integrated matrix elements (\ref{twobody}) and in addition one 
can easily read off the isospin of the corresponding wave function.


For the irrep $[22]$  there are two basis vectors and their expressions
are straightforward because the pair 12 and 34 are always either in a symmetric or antisymmetric pair. 
We have  
\begin{eqnarray}\label{22sprime}
|[22]_F 2211 \rangle &=& \frac{1}{2} [ \phi_{[2]}(us)~\phi_{[2]}(cd)  +  \phi_{[2]}(cd)~\phi_{[2]}(us)\nonumber \\ 
& - & \phi_{[2]}(sd)~\phi_{[2]}(uc)  -  \phi_{[2]}(uc)~\phi_{[2]}(sd) ]
\end{eqnarray}
and 
\begin{eqnarray}\label{22aprime}
|[22]_F 2121 \rangle  &=& \sqrt{\frac{1}{12}}[2 \phi_{[11]}(ud)~\phi_{[11]}(sc)
 +   2 \phi_{[11]}(sc)~\phi_{[11]}(ud)\nonumber \\
& + & \phi_{[11]}(uc)~\phi_{[11]}(sd) + \phi_{[11]}(sd)~\phi_{[11]}(uc)\nonumber \\
& - & \phi_{[11]}(us)~\phi_{[11]}(cd) - \phi_{[11]}(cd)~\phi_{[11]}(us)]
\end{eqnarray}
where (\ref{22aprime}) obviously has isospin $I = 0$
which means that the pairs 12 and 34  in (\ref{22sprime})  have to couple to the same isospin as well.


For irrep $[31]_F$ the vectors $[31]_{F_1}$ and $[31]_{F_2}$ have to be combined
in the so called Young-Yamanouchi-Rutherford basis first proposed in the
context of nuclear physics \cite{Harvey:1988nk,Harvey:1980rva}. It is defined such as 
the last two particles are either in a symmetric or an antisymmetric state  
The pair 12 is also in a symmetric or an antisymmetric state, which is very advantageous. For more than four particles 
the problem is more complicated. Here we have \cite{Stancu:1991rc}
\begin{eqnarray}\label{31s}
  | [31]_F {\overline {12}11} \rangle  &=& \sqrt{\frac{2}{3}}~| [31]_F {1211} \rangle 
+ \sqrt{\frac{1}{3}}~| [31]_F  {2111} \rangle 
\end{eqnarray}
where in the left hand side both pairs 12 and 34 are in a symmetric state 
and 
\begin{eqnarray}\label{31a}
  | [31]_F { \tilde {12}11} \rangle  &=& \sqrt{\frac{1}{3}}~| [31]_F {1211} \rangle 
- \sqrt{\frac{2}{3}}~| [31]_F  {2111} \rangle 
\end{eqnarray}
where the pair 12 is in a symmetric and 34  in an antisymmetric state.
Using 
Eqs. (A.16) and  (A.15) of \cite{Yuan:2012wz}, defining $[31]_{F_2}$ and   $[31]_{F_1}$ respectively,
one obtains
\begin{eqnarray}\label{31sprime} 
  | [31]_F {\overline {12}11} \rangle  &=& \frac{1}{2} [\phi_{[2]}(us)~\phi_{[2]}(cd) - \phi_{[2]}(cd)~\phi_{[2]}(us) \nonumber \\
& + & \phi_{[2]}(uc)~\phi_{[2]}(ds) - \phi_{[2]}(ds)~\phi_{[2]}(uc)],
\end{eqnarray}
and
\begin{eqnarray}\label{31aprime}
| [31]_F {\tilde {12}11} \rangle  &=&\sqrt{\frac{1}{8}}[\phi_{[2]}(uc)~\phi_{[11]}(ds) - \phi_{[2]}(us)~\phi_{[11]}(cd)  \nonumber \\
& - & \phi_{[2]}(cd)~\phi_{[11]}(us) - \phi_{[2]}(ds)~\phi_{[11]}(uc)  \nonumber \\
& - & 2 \phi_{[2]}(sc)~\phi_{[11]}(ud)],
\end{eqnarray}. 
The state (\ref{31aprime}) obviously has I = 0 thus  (\ref{31sprime}) should also have I = 0.

The third basis vector  $[31]_{F_3}$ of Ref. \cite{Yuan:2012wz} can simply be rewritten as 
\begin{eqnarray}\label{31saprime}
| [31]_F 1121 \rangle  &=&\sqrt{\frac{1}{8}}[2 \phi_{[11]}(ud)~\phi_{[2]}(sc) - \phi_{[11]}(ds)~\phi_{[2]}(uc) + \phi_{[11]}(cd)~\phi_{[2]}(us)\nonumber \\
& + & \phi_{[11]}(us)~\phi_{[2]}(cd) + \phi_{[11]}(uc)~\phi_{[2]}(ds)] ,
\end{eqnarray}
where the pair 12 is in an antisymmetric state and 34 in a symmetric state. The state obviously has I = 0. 


\begin{table*}
\caption{\label{flavorstates} The $I$ = 0 udsc flavor states in two different notations and the corresponding Weyl tableaux.}
\begin{tabular}{ccccccc}
\hline
Young-Yamanouchi & \hspace{2mm} Ref. \cite{Yuan:2012wz} & \hspace{2mm} Weyl tableau \\[1.1ex]
\hline\\
$ [22]_F 2211 $ & \hspace{1mm} $[22]_{F_1}$ & \hspace{2mm} 
\raisebox{-9.0pt}{\mbox{\begin{Young}
u & s  \cr 
d & c \cr
\end{Young}}}\\[1.1ex]
$ [22]_F 2121 $ & \hspace{1mm} $[22]_{F_2}$ & \hspace{2mm} \\[1.1ex]
\hline 
& & & \\
$ [31]_F 2111 $ & \hspace{1mm} $[31]_{F_1}$ & \hspace{2mm} 
\raisebox{-9.0pt}{\mbox{\begin{Young}
u & s & c \cr
d \cr
\end{Young}}}\\[1.1ex]
$ [31]_F 1211 $ & \hspace{1mm} $[31]_{F_2}$ & \hspace{2mm} \\[1.1ex]
$ [31]_F 1121 $ & \hspace{1mm} $[31]_{F_3}$ & \hspace{2mm} \\[1.1ex]
\hline
& & & \\
$ [211]_F 3211 $ & \hspace{1mm} $[211]_{F_1}$ & \hspace{2mm} 
\raisebox{-9.0pt}{\mbox{\begin{Young}
u &  s \cr
d \cr
c \cr
\end{Young}}}\\[1.1ex]
$ [211]_F 3121 $ & \hspace{1mm} $[211]_{F_2}$ & \hspace{2mm} \\[1.1ex]
$ [211]_F 1321 $ & \hspace{1mm} $[211]_{F_3}$ & \hspace{2mm} \\[1.1ex]
\hline
& & & \\
$ [211]^{'}_F 3211 $ & \hspace{1mm} $[211]^{'}_{F_1}$ & \hspace{2mm} 
\raisebox{-9.0pt}{\mbox{\begin{Young}
u &  c \cr
d \cr
s \cr
\end{Young}}}\\[1.1ex]
$ [211]^{'}_F 3121 $ & \hspace{1mm} $[211]^{'}_{F_2}$ & \hspace{2mm} \\[1.1ex]
$ [211]{'}_F 1321 $ & \hspace{1mm} $[211]^{'}_{F_3}$ & \hspace{2mm} \\[1.1ex]
\hline
\end{tabular}
\end{table*}

For the irrep $[211]_F$ 
the Young-Yamanouchi-Rutherford basis vectors are 
\begin{eqnarray}\label{211s}
  | [211]_F {\overline {13}21} \rangle  &=& \sqrt{\frac{2}{3}}~| [211]_F {1321} \rangle 
+ \sqrt{\frac{1}{3}}~| [211]_F  {3121} \rangle 
\end{eqnarray}
where the pair 12 is in an antisymmetric and 34  in a symmetric state and
\begin{eqnarray}\label{211a}
  | [211]_F { \tilde {13}21} \rangle  &=& \sqrt{\frac{1}{3}}~| [211]_F {1321} \rangle 
- \sqrt{\frac{2}{3}}~| [31]_F  {3121} \rangle 
\end{eqnarray}
where both pairs 12 and 34  are in an antisymmetric state. Using Eqs. (A.20) and (A.19) of 
Ref. \cite{Yuan:2012wz} one obtains
\begin{eqnarray}\label{211sprime}
  | [211]_F {\overline {13}21} \rangle  &=& \sqrt{\frac{1}{24}} [ 2~\phi_{[11]}(ud)~\phi_{[2]}(sc) 
-  3 \phi_{[11]}(uc)~\phi_{[2]}(ds) - 3 \phi_{[11]}(cd)~\phi_{[2]}(us)  \nonumber \\
& + &   \phi_{[11]}(us)~\phi_{[2]}(cd) - \phi_{[11]}(ds)~\phi_{[2]}(uc)] 
\end{eqnarray}
and
\begin{eqnarray}\label{211aprime}
  | [211]_F { \tilde {13}21} \rangle  &=& \sqrt{\frac{1}{12}} ~[ - 2 \phi_{[11]}(ud)~\phi_{[11]}(sc) + 2 \phi_{[11]}(sc)~\phi_{[11]}(ud)
-  \phi_{[11]}(uc)~\phi_{[11]}(ds)    \nonumber \\
& - &  \phi_{[11]}(cd)~\phi_{[11]}(us) + \phi_{[11]}(us)~\phi_{[11]}(cd) + \phi_{[11]}(ds)~\phi_{[11]}(uc)].
\end{eqnarray}

The vector $[211]_{F_1}$  of Ref. \cite{Yuan:2012wz} can be rewritten as
\begin{eqnarray}\label{211sameprime}  
| [211]_F 3211 \rangle  &=&  \sqrt{\frac{1}{24}} [ \phi_{[2]}(uc)~\phi_{[11]}(sd) +  \phi_{[2]}(cd)~\phi_{[11]}(us)  
+ 2 \phi_{[2]}(cs)~\phi_{[11]}(ud) \nonumber \\
& - & 3 \phi_{[2]}(sd)~\phi_{[11]}(uc) - 3 \phi_{[2]}(us)~\phi_{[11]}(cd)].
\end{eqnarray}
For the irrep $[211]^{'}_F$ the Young-Yamanouchi-Rutherford basis vectors are
defined like in  Eqs.   (\ref{211s}) and  (\ref{211a}) but in  the right hand side
one must use  the vectors $[211]^{'}_{F_i}$ instead of  $[211]_{F_i}$, i. e. Eqs. (A.23) and (A.22)
of  Ref. \cite{Yuan:2012wz}.
One obtains
\begin{eqnarray}\label{211primesprime} 
  | [211]^{'}_F {\overline {13}21} \rangle  &=& \sqrt{\frac{1}{3}}[\phi_{[11]}(ud)~\phi_{[2]}(sc) - \phi_{[11]}(us)~\phi_{[2]}(cd)\nonumber \\
& + & \phi_{[11]}(ds)~\phi_{[2]}(uc) ],
\end{eqnarray}
and
\begin{eqnarray}\label{211primeaprime}
| [211]^{'}_F {\tilde {13}21} \rangle  &=&\sqrt{\frac{1}{6}}[ \phi_{[11]}(ud)~\phi_{[11]}(sc) + \phi_{[11]}(us)~\phi_{[11]}(cd)
+ \phi_{[11]}(ds)~\phi_{[11]}(uc)\nonumber \\
& - & \phi_{[11]}(cd)~\phi_{[11]}(us) - \phi_{[11]}(sc)~\phi_{[11]}(ud) -  \phi_{[11]}(uc)~\phi_{[11]}(ds)].
\end{eqnarray}
They obviously have I = 0.
The third basis vector $[211]^{'}_{F_1}$ can be rewritten in the convenient form
\begin{eqnarray}\label{211primesaprime}
| [211]^{'}_F { 3211} \rangle  &=& \sqrt{\frac{1}{3}}[ \phi_{[2]}(sc)~\phi_{[11]}(ud) - \phi_{[2]}(cd)~\phi_{[11]}(us)\nonumber \\
& - & \phi_{[2]}(uc)~\phi_{[11]}(sd) ]
\end{eqnarray}
which also has I = 0.


\vspace{1cm}

\acknowledgements

This work has been supported by the Fonds de la Recherche Scientifique - FNRS, Belgium, 
under the grant number 4.4503.19.

\end{document}